\pdfoutput=1
\documentclass[sigconf]{acmart}  
\usepackage{cite}
\usepackage{amsmath}
\usepackage{enumerate}
\usepackage{makecell}
\usepackage{CJKutf8}  
\usepackage{stfloats}
\usepackage{etoolbox}
\usepackage[ruled,linesnumbered]{algorithm2e}
\usepackage{graphicx}
\usepackage{multirow}  
\usepackage{xcolor}  
\usepackage{hyperref}
\usepackage{booktabs}  
\definecolor{MyLightGreen}{RGB}{230,255,204}
\definecolor{MyLightBlue}{RGB}{204,229,255}
\definecolor{MyLightGrey}{RGB}{230,230,230}
\definecolor{MyLime}{RGB}{102,255,102}
\definecolor{MyLightCoral}{RGB}{255,153,153}

\renewcommand\footnotetextcopyrightpermission[1]{}
\AtBeginDocument{%
\providecommand\BibTeX{{%
\normalfont B\kern-0.5em{\scshape i\kern-0.25em b}\kern-0.8em\TeX}}}

\setcopyright{acmcopyright}
\copyrightyear{2023}
\acmYear{2023}
\acmDOI{XXXXXXX.XXXXXXX}

\acmPrice{15.00}
\acmISBN{978-1-4503-XXXX-X/18/06}

\title{Event-driven Real-time Retrieval in Web Search}

\author{Nan Yang}
\affiliation{%
    \institution{Tencent PCG}
    \city{Beijing}
    \country{China}}
\email{marinyang@tencent.com}

\author{Shusen Zhang}
\affiliation{%
    \institution{Tencent PCG}
    \city{Beijing}
    \country{China}}
\email{shusenzhang@tencent.com}

\author{Yannan Zhang}
\affiliation{%
    \institution{Tencent PCG}
    \city{Beijing}
    \country{China}}
\email{yananzhang@tencent.com}

\author{Xiaoling Bai}
\affiliation{%
    \institution{Tencent PCG}
    \city{Beijing}
    \country{China}}
\email{devinbai@tencent.com}

\author{Hualong Deng}
\affiliation{%
    \institution{Tencent PCG}
    \city{Beijing}
    \country{China}}
\email{tonnydeng@tencent.com}

\author{Tianhua Zhou}
\affiliation{%
    \institution{Tencent PCG}
    \city{Beijing}
    \country{China}}
\email{kivizhou@tencent.com}

\author{Jin Ma}
\affiliation{%
    \institution{USTC}
    \city{Hefei}
    \country{China}}
\email{majin01@mail.ustc.edu.cn}

\vspace{2em}
\begin{document}

\begin{abstract}

Information retrieval in real-time search presents unique challenges distinct from those encountered in classical web search. These challenges are particularly pronounced due to the rapid change of user search intent,
which is influenced by the occurrence and evolution of breaking news events, such as earthquakes, elections, and wars.
Previous dense retrieval methods, which primarily focused on static semantic representation, lack the capacity to capture immediate search intent, leading to inferior performance in retrieving the most recent event-related documents in time-sensitive scenarios.
To address this issue, this paper expands the query with event information that represents real-time search intent.
The Event information is then integrated with the query through a cross-attention mechanism, resulting in a time-context query representation. We further enhance the model's capacity for event representation through multi-task training.
Since publicly available datasets such as MS-MARCO do not contain any event information on the query side and have few time-sensitive queries,
we design an automatic data collection and annotation pipeline to address this issue, which includes ModelZoo-based Coarse Annotation and LLM-driven Fine Annotation processes.
In addition, we share the training tricks such as two-stage training and hard negative sampling.
Finally, we conduct a set of offline experiments on a million-scale production dataset to evaluate our approach and deploy an A/B testing in a real online system to verify the performance.
Extensive experimental results demonstrate that our proposed approach significantly outperforms existing state-of-the-art baseline methods.

\end{abstract}

\vspace{-1ex}

\ccsdesc[300]{Information systems~Retrieval models and ranking}

\vspace{-1ex}
\keywords{Information retrieval, Real-time search, Large Language Model}



\maketitle


\vspace{-1.5ex}
\section{Introduction}
\label{sec:introduction}

\begin{figure}[thb]
    \centering
    \includegraphics[scale=0.85]{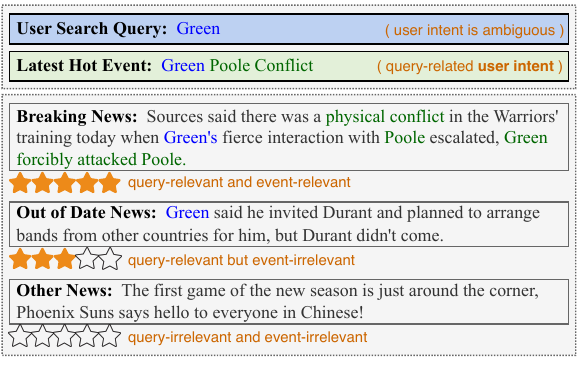}
    \vspace{-1em}
    \caption{In the realm of time-sensitive search scenarios, given a user search query, the most likely query intent is defined as the latest trending event related to the query.
    Consequently, we can categorize documents into three tiers, ranging from high to low quality:
    1) Breaking news that is both query-relevant and event-relevant;
    2) The out-of-date news that is query-relevant but event-irrelevant;
    3) Other news that is neither query-relevant nor event-relevant.
    }
    \label{fig:case_demo}
    \vspace{-1em}
\end{figure}

Over the past decades, news search has become an increasingly important portal for people to access information.
As an important component of news search, real-time retrieval~\citep{real_time_search} has emerged as a critical requirement, as it places greater emphasis on the timeliness of retrieved documents compared to traditional dense retrieval methods.
The fundamental challenge in information retrieval lies in calculating the similarity between a query and a document, which can be achieved through literal matching or semantic matching. While traditional methods like BM25\citep{robertson2009probabilistic} are effective for literal matching, they fall short in semantic matching. To address this issue, large-scale pre-trained models have been successfully employed for semantic retrieval ~\citep{huang2020embedding, khattab2020colbert,humeau2020polyencoders, lu2022erniesearch, liu2021pretrained, liu2021que2search}. However, real-time retrieval poses unique challenges and characteristics in our specific context:

On the one hand, real search intent changes rapidly with the occurrence and evolution of breaking news.
The query representation encoded by pre-trained language models~(PTMs) is a static vector that does not contain any requirements corresponding to the current event.
Due to the lack of real-time context, event-aware documents can not be adopted, especially for short and long-tail queries.
As shown in Figure~\ref{fig:case_demo},
in news search, when users enter the query ``Green'',
they are highly likely trying to find the breaking news, e.g. ``Green Poole Conflict''. Unfortunately, the intent of the original query is ambiguous and there are no differences in the semantic scores between \textit{event-relevant} and \textit{event-irrelevant} documents.
Therefore, the event-relevant documents may be ranked lower or truncated, making it difficult to meet the user intent.

On the other hand, existing retrieval benchmarks, such as MS-MARCO~\citep{nguyen2016ms}, predominantly concentrate on general search scenarios, which have a different data distribution from time-sensitive queries. Additionally, traditional datasets are usually constructed by mining based on click signals or manual annotations. Nevertheless, the click-based approach is unsuitable for news search due to the sparsity of user click data. Simultaneously, manual annotation proves to be both inefficient and costly. Therefore, there is an urgent need for a fast, efficient, and low-cost data annotation method specifically tailored to time-sensitive search scenarios.

To tackle the unique challenges in real-time retrieval, we propose a novel approach called \textbf{E}vent-driven \textbf{R}eal-time \textbf{R}etieval~(\textbf{ERR}) in this paper. ERR mainly focuses on the following aspects:
1) We introduce a new two-tower model that optimizes retrieval performance by focusing on query event expansion.
For time-sensitive queries, accurately describing the latest query intent is crucial. To achieve this, we use event-centric query expansion~\citep{Zhang2023EventCentricQE} to obtain real-time events related to the query and extend the query intent by fusing query and hot event information. Events effectively help retrieve more timely documents by providing supplementary information for queries. In this study, we effectively use Adaptive Cross-Attention~\citep{cross_attention} and MT-DNN~\citep{liu2019multi} for event data fusion. Cross-Attention is widely used in natural language understanding (e.g., Transformer\citep{vaswani2017attention}) to fuse multiple texts and in computer vision (e.g., CrossVit\citep{chen2021crossvit}) to fuse different modal data. Additionally, multi-task training is used to make the model more focused on event information.
2) To effectively obtain data for timely retrieval and reduce data annotation costs, we propose a two-stage automatic data annotation approach consisting of a \textit{ModelZoo-based Coarse Annotation} and an \textit{LLM-driven Fine Annotation}. Firstly, we collected a large amount of unsupervised data and used multiple models for majority voting, to mine easy samples with high confidence. In the second stage, we further utilized the powerful semantic understanding ability of large language models~(LLMs) to perform fine-grained annotation on the uncertain voting results from the first stage.
We conducted a thorough investigation and comparison of various instructions to achieve more accurate data annotation outcomes.
Our method has been successfully deployed to an online retrieval system. Numerous offline and online experiments have demonstrated that ERR dramatically improves the performance of real-time retrieval.

To highlight, this paper proposes a novel retrieval approach called ERR, which contributes mainly to the following aspects:
\vspace{-0.15ex}
\begin{itemize}
\item We propose a novel real-time retrieval model that fuses events and queries through a cross-attention and multi-task mechanism to recall more real-time documents.
\item To obtain data effectively and reduce data annotation costs for real-time retrieval, we introduce a two-stage automatic sample annotation pipeline consisting of a ModelZoo-based Coarse Annotation and an LLM-driven Fine Annotation.
\item We conduct numerous offline and online experiments that demonstrate the superiority of ERR over existing state-of-the-art models in real-time retrieval tasks.
\end{itemize}


\vspace{-2.5ex}
\section{related work}

\subsection{Information retrieval}

Information retrieval aims to provide users with the information they need, focusing on evaluating the correlation between a query and a document. Methods can be categorized into traditional retrieval models and neural network retrieval models. Traditional models, like BM25~\citep{robertson2009probabilistic}, rely on accurate matching signals but often fall short in semantic matching as they primarily consider literal matching.
Neural network models are widely employed in information retrieval. DSSM~\citep{huang2013learning} learns feature representations for queries and documents, calculating correlation scores through inner product. ARC-I~\citep{hu2014convolutional} and CLSM~\citep{shen2014latent} utilize CNN to capture word order and context information. LSTM-RNN~\citep{palangi2016deep} enhances query and document representations using LSTM. NRM-F~\citep{zamani2018neural} achieves good performance by considering document content, title, and other contents at the coding level.
Pre-training technology has gained attention in deep learning, leading to various strategies in information retrieval. Models like BERT~\citep{devlin2018bert} and ERNIE~\citep{sun2019ernie}, built on pre-training, greatly enhance representation ability for queries and documents. Sentence embedding, used in retrieval, matching, and classification, is improved by models like Sentence-BERT~\citep{reimers2019sentence}, employing Siamese and triplet networks. Contrastive learning methods such as SimCSE~\citep{gao2021simcse}, have also achieved success in semantic similarity retrieval.

\vspace{-0.8em}
\subsection{LLM-Driven Data Annotation}

LLMs gain significant attention due to their exceptional performance across various natural language processing tasks, with the flourishing development of ChatGPT~\citep{chatgpt}, GPT-4~\citep{gpt4} and LLaMA~\citep{touvron2023llama}.
A growing number of studies showcase LLM-driven data annotation potential in various language tasks, highlighting its effectiveness and promising prospects for diverse applications. \citet{kim2023cotever} introduced a toolkit for annotating factual correctness in chain-of-thought~(CoT) prompting, addressing factuality challenges and enhancing faithfulness. \citet{Data-Copilot} proposed an LLM-based system for autonomously managing, processing, and displaying heterogeneous data, serving as a reliable AI assistant in diverse industries. \citet{kuzman2023chatgpt} utilized document embeddings with ChatGPT or GPT-4 for text annotations, achieving competitive performance in text classification, sentiment analysis, and topic modeling. \citet{yu2023using} found ChatGPT surpassed a fine-tuned multilingual XLM-RoBERTa model in automatic genre identification on an unseen dataset, with native speakers evaluating generated examples in different languages. In-context learning capabilities of LLMs were explored through an annotation-efficient, two-step framework for new language tasks~\citep{Selective_Annotation}, where the unsupervised, graph-based selective annotation method, vote-k, significantly improved performance and reduced annotation costs compared to supervised fine-tuning approaches.

\begin{figure*}[ht]
    \centering
    \includegraphics[scale=0.65]{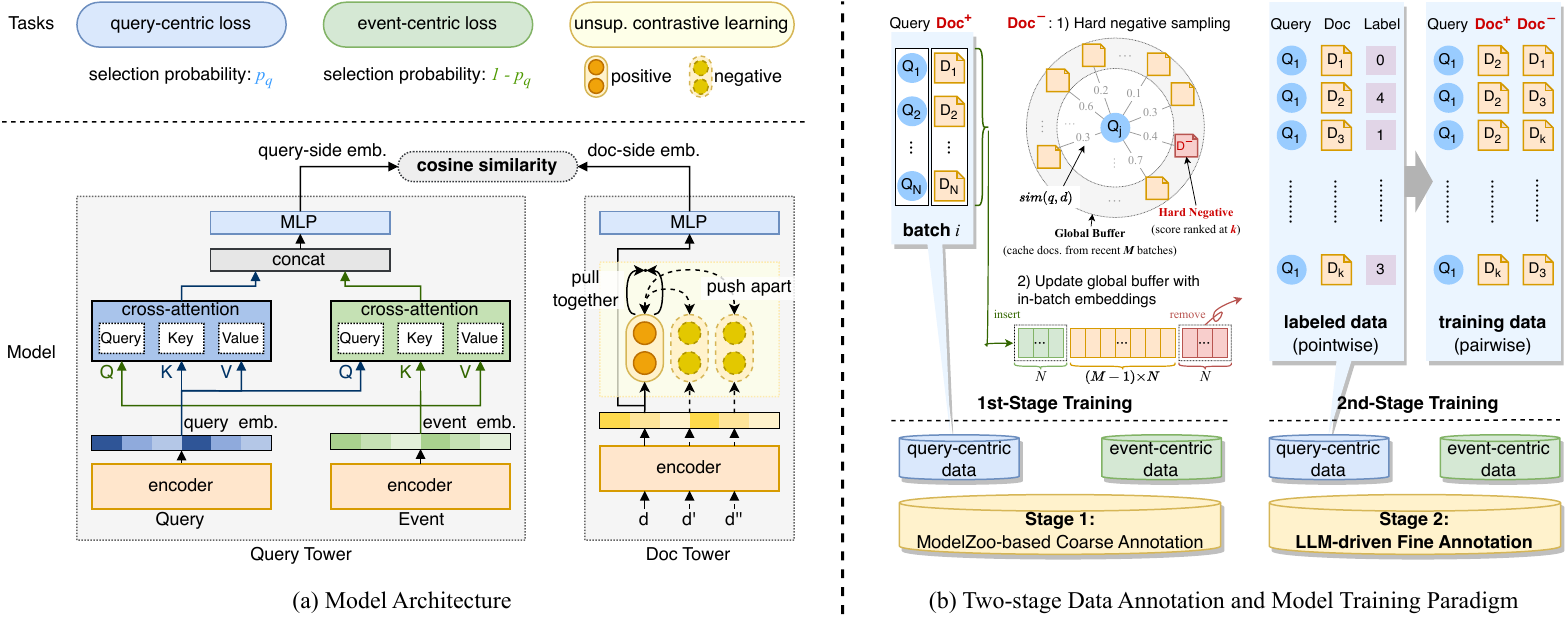}
    \vspace{-1em}
    \caption{Method Overview.}
    \label{fig:model_arch}
\end{figure*}

\vspace{-1ex}
\section{methodology}

In this section, we provide a detailed introduction to the various aspects of ERR, including the retrieval model and data annotation components. As shown in the figure~\ref{fig:model_arch}, the model has several aspects to consider. In the query-end, we incorporate event information into the real-time search intent of the query and fuse them together using a cross-attention mechanism~($\S$~\ref{sec:EventAug}, $\S$~\ref{sec:EventFu}). In the document-end, unsupervised contrastive learning is leveraged to augment the capacity for representing textual semantics~($\S$~\ref{sec:TO}). The training data is categorized into two types - query-centric samples and event-centric samples. During the training phase, both objectives are optimized simultaneously in a multi-task manner~($\S$~\ref{sec:MultiT}). In terms of data annotation, a two-stage approach is proposed, comprising a ModelZoo-based coarse annotation and an LLM-driven fine annotation~($\S$~\ref{sec:DataCa}).

\vspace{-1ex}
\subsection{Event Augment} \label{sec:EventAug}

\begin{figure}[t]
    \centering
    \includegraphics[scale=0.6]{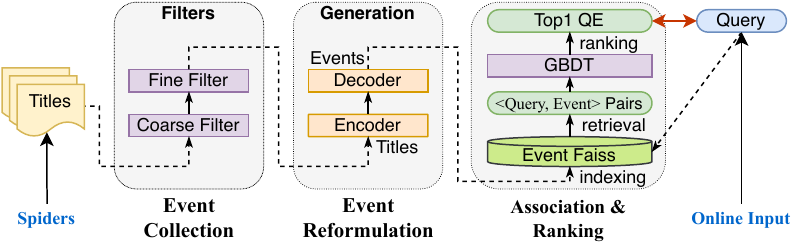}
    \vspace{-1em}
    \caption{Illustration of the event augment process.}
    \vspace{-1.5em}
    \label{fig:event_from}
\end{figure}

We draw inspiration from the approach proposed in~\citep{Zhang2023EventCentricQE} to identify and select the most fulfilling event as a query expansion. As shown in figure~\ref{fig:event_from}, the methodology consists of the following steps:

\begin{enumerate}
  \item \textbf{Event Collection:} Gathering a stream of event titles from various sources and performing rule-based coarse filtering followed by semantic-based fine filtering to obtain event candidates.

  \item \textbf{Event Reformulation:} Using a generated model to analyze the collected event titles, extract key information from them, and discard noise information.

  \item \textbf{Event Association:} By utilizing semantic retrieval techniques, specifically with the help of faiss~\citep{johnson2019billion}, we establish associations between queries and events, allowing for a deeper understanding of their relationships.

  \item \textbf{Online Ranking:} Integrating additional features, such as event found time and event popularity~(the size of the cluster to which an event belongs), into the event candidates, not just relying on relevance alone, and applying GBDT~\citep{friedman2001greedy} as a ranking model to establish a more accurate matching relationship between the events and the query.
\end{enumerate}

By following this systematic approach, we choose the event candidate with the highest score as the query expansion.

\subsection{Event Fusion}

We use the event as a supplement to the original query, and both it and the original query participate in the search, to obtain richer and more accurate matching documents.

\subsubsection{Cross-Attention}\label{sec:EventFu}
To make better use of event information and to retain crucial information from the original search at the same time,
we use Adaptive Cross-Attention~\citep{cross_attention} to fuse these two domains.
E.g.
In cases where the event and query have weak relevance, the embedding of the query tower may lean more towards the semantic representation of the original user search query.

Given a query $q_1$ and an event $q_2$ ,
we utilize PTM like BERT to encode them and get their embedding representations respectively,
and then fuse the semantic information of the two segments by cross-attention to get the new embedding $\mathbf{CA}_{i} \in \mathbb{R}^{1 \times d}$.
Mathematically, the $\mathbf{CA}$ can be expressed as
\begin{equation}
    \begin{gathered}
        \mathbf{Q} = {\mathbf{x}^l_{q_j}}\mathbf{W}_Q, \ \ \
        \mathbf{K} = \mathbf{x}^l_{q_i}\mathbf{W}_K, \ \ \
        \mathbf{V} = \mathbf{x}^l_{q_i}\mathbf{W}_V \ \\
        \mathbf{CA_i} = \mathrm{softmax}(\frac{\mathbf{Q_j}\mathbf{K_i}^T}{\sqrt{C / h}}) \mathbf{V_i}
        \label{eq:CA}
    \end{gathered}
\end{equation}
where $i,j=1,2; i \ne j$ denote different input data indexes, i.e. query or event.
$\mathbf{W}_Q$, $\mathbf{W}_K$, $\mathbf{W}_V$ $\in \mathbb{R}^{C\times (C/h)}$ are learnable parameters,
$\mathbf{x}^l_{q_1}, \mathbf{x}^l_{q_2} \in \mathbb{R}^{L\times C}$, $L$, $C$ and $h$ denote the number of words in each sentence, the embedding dimension and number of heads, respectively.

Besides the cross-attention, each query or event tower also contains a fully connected feed-forward network that is applied to each position separately and identically.
The feed-forward network consists of two linear transformations with an activation function ReLU in between.
The last hidden layer of the BERT encoder is fed into a cross-attention based transformer block and obtains the final representation.
The process mentioned above can be written as:
\begin{gather}
    \mathrm{Trm} = {\operatorname{max}}(0, x\mathbf{W_1} + b_1)\mathbf{W_2} + b_2
    \label{eq:cross_attention_transformer}
\end{gather}
where $x$ is the cross-attention layer, $\mathbf{W}_1$, $\mathbf{W}_2$, $b_1$, $b_2$ are learnable parameters.

To better represent the fused embedding, the transformer outputs of query and event are concatenated and then applied to a multi-layer perceptron.
Formally, the query side semantic representation $q_{emb}$ is obtained as follows,
\begin{gather}
    q_{emb} = \mathrm{MLP}(\mathrm{Trm}(query) \oplus \mathrm{Trm}(event))
    \label{eq:q_emb}
\end{gather}
where $\oplus$ represents the concatenate operation.

Considering the difference in the distribution of session queries between online and offline, we use an adaptive approach to fuse event and query information to solve the problem of low event coverage.
In the case of missing event fields, we use the query itself to complement the event fields, which means that $\mathbf{x}^l_{q_1}$, $\mathbf{x}^l_{q_2}$ are equivalent.
With this treatment, the model structure remains consistent even in cases of missing events, and the training time is reduced.

\subsubsection{Multi-Task Training}\label{sec:MultiT}
To make the model more focused on event information, we also introduce multi-task training to our approach.
The dataset $\mathcal{D}$ which containing $K$ training examples is defined as follows,
\begin{equation}
    \mathcal{D} = \left\{ \left(q_i, e_i, d^+_i, d^-_i \right) \right\}^{K}_{i=1}
    \label{eq:dataset}
\end{equation}
where each training example is a quadruplet composed of:
a query $q_i$,
an event $e_i$ that related to the query $q_i$,
a positive document $d_i^+$,
and a negative document $d_i^-$.

We divide the training data into two kinds of datasets:
The first type is query-centric samples: $\mathcal{D}_q = \left\{(q, e, d^+, d^-)\right\}$, in which all the positive documents are query-relevant and are possibly event-irrelevant, denoted as $r(q, d^+)=1, r(e, d^+)=0\ or \ 1$.
Since the default premise of our task is that each event is related to the query, the documents which are irrelevant to the query are absolutely irrelevant to its corresponding event,
we denote it as $r(q, d^-)=0, r(e, d^-)=0$.

In contrast, the second type is event-centric samples: $\mathcal{D}_e=\left\{(q, e, d^+, d^-)\right\}$, which means all the positive documents are event-relevant as well as query-relevant, we express it as $r(q, d^+)=1, r(e, d^+)=1$.
As for negative documents, they are event-irrelevant and potentially query-irrelevant, denoted as $r(e, d^-)=0, d(q, d^-)=0\ or\ 1$.

Both query-centric samples and event-centric samples employ triplet loss with margin $\delta$:
\begin{equation}
\begin{gathered}
    \small\mathcal{L}(\mathcal{D}_q) = \sum_{(q_i, e_i, d^+_i, d^-_i) \in \mathcal{D}_q} \max\left(0, \delta - f(q_i, e_i, d^+_i) + f(q_i, e_i, d^-_i)\right)
    \\
    \small\mathcal{L}(\mathcal{D}_e) = \sum_{(q_i, e_i, d^+_i, d^-_i) \in \mathcal{D}_e} \max\left(0, \delta - f(q_i, e_i, d^+_i) + f(q_i, e_i, d^-_i)\right)
\end{gathered}
\label{eq:triplet_loss}
\end{equation}
where $\small\mathcal{L}(\mathcal{D}_q), \small\mathcal{L}(\mathcal{D}_e)$ can be considered as the objective of the query-centric task and event-centric task, respectively.

We apply MT-DNN training algorithm\citep{liu2019multi} to train our model.
In the training stage, the training data in each mini-batch is randomly selected from one of the aforementioned samples with the probability of $p_t$,
and the model is updated according to the task-specific objective for the task $t$.
The overall task optimization objective thus can be expressed as:
\begin{equation}
    \mathcal{L}_{t} = \begin{cases} \mathcal{L}(\mathcal{D}_q) & x > p_q \\ \mathcal{L}(\mathcal{D}_e) & otherwise \end{cases}
\label{eq:task_objective}
\end{equation}
where $x \sim \mathcal{U}(0, 1)$ is a random number following uniform distribution in the range of $[0,1]$, $p_q$ is the pre-defined probability of the query-centric task.

\subsection{Optimization Objective}
\label{sec:TO}
To enhance the model's capability to characterize unknown documents during training, we introduce unsupervised contrastive learning to the document tower.
We denote $\mathbf{h}_i^z = {f_{\theta}({x_i}, z)}$, where $z$ is a random mask for dropout, ${x_i}$ is the sentence in our dataset.
We simply feed the same input to the encoder twice to obtain two [CLS] embeddings $\mathbf{h}_i, \mathbf{h}_{i}^{+}$ with different dropout masks ${z}$ and ${z'}$, $\mathbf{h}_i$ and $\mathbf{h}_i^+$ are semantically close.
We regard $\mathbf{h}_i^+$ as positive of $\mathbf{h}_i$ and other sentences' embedding in the same mini-batch as negatives.
Then the training objective of unsupervised contrast learning becomes:
\begin{equation}
    \label{eq:unsup_contra}
    \begin{aligned}
        \mathcal{L}_{CL}=-\log \frac{e^{\operatorname{sim}\left(\mathbf{h}_i, \mathbf{h}_{i}^{+} \right) / \tau}}{\sum_{j=1}^N e^{\operatorname{sim}\left(\mathbf{h}_i, \mathbf{h}_{j}^{+}\right) / \tau}}
    \end{aligned}
\end{equation}
where $\tau$ is a temperature hyper-parameter, $N$ is the mini-batch size.

The final training objective is a linear combination of the triplet task-specific loss and the unsupervised contrast loss:
\vspace{-0.05in}
\begin{equation}
    \label{eq:total_loss}
    \mathcal{L} = \mathcal{L}_t + \lambda \cdot \mathcal{L}_{CL}
\end{equation}
where $\small\mathcal{L}_t$ is the task loss defined in Eq. (\ref{eq:task_objective}),
$\mathcal{L}_{CL}$ is the unsupervised contrastive learning loss defined in Eq. (\ref{eq:unsup_contra});
$\lambda$ is a hyper-parameter controlling the trade-off between $\small\mathcal{L}_t$ and $\mathcal{L}_{CL}$.

\subsection{Data Collection And Annotation}\label{sec:DataCa}
In the model training stage,
real-time retrieval faces the following problems:
1) Existing public datasets, such as MS-MARCO~\citep{nguyen2016ms}, do not contain any event information on the query side.
Besides, these datasets mostly focus on general search scenarios, which have significant differences in data distribution from time-sensitive scenarios, e.g. too few time-sensitive queries are included.
2) Traditional methods such as~\citep{huang2020embedding},~\citep{zou2021pre} adopt user clicks as the relevance label.
Unfortunately,
compared with classical web search, there are more newly published news documents in real-time search results,
resulting in sparse click data and significant data noise, especially for negative samples.

In addition, human annotation proves to be both inefficient and costly.
Therefore, we collect authentic data from the production environment and annotate it using our automated annotation pipeline, which we will discuss in detail in sections~\ref{subsec:collection} and~\ref{subsec:annotation}.

\begin{figure*}[ht]
    \centering
    \includegraphics[scale=0.75]{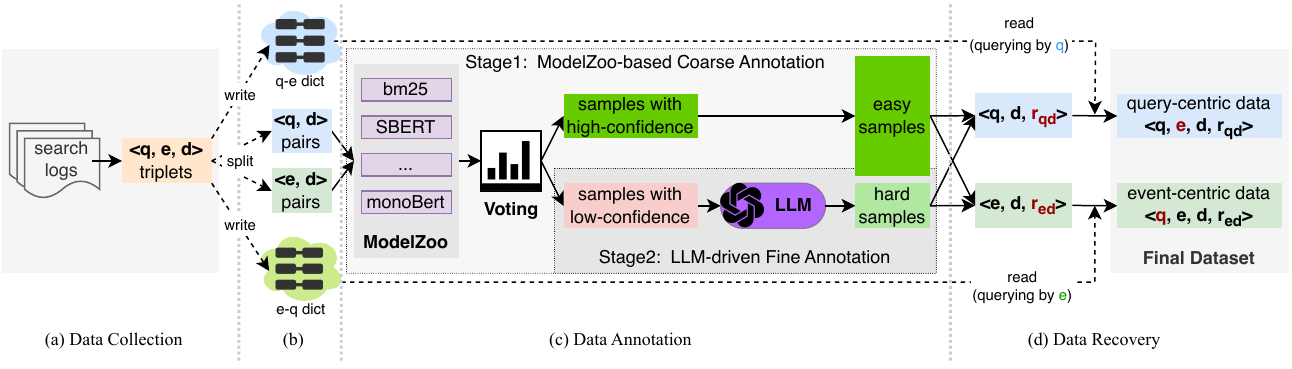}
    \vspace{-1em}
    \caption{
        Depiction of our data collection and annotation process.
    }
    \label{fig:data_arch}
\end{figure*}

\subsubsection{Data Collection}
\label{subsec:collection}
Both training data and testing data are collected from the real production environment.

\textbf{Training Data}
The training data is randomly derived from the search logs in two consecutive months, consisting of the following parts:
1) The query input by the user. 2) Event information related to the query. 3) Corresponding documents in the search results.
We denote each sample as a $\left<q, e, d\right>$ triplet.
We filter out samples whose queries do not exhibit a real-time search intent before annotation.
The data is annotated through our automatic data annotation pipeline, we will describe more details in section~\ref{subsec:annotation}.

\textbf{Testing Data}
\label{subsec:test_data}
The testing data shares similar elements and distribution with the training data but is collected from search logs in different time periods to prevent information leakage.
We annotated the testing data using our crowd-sourcing platform,
where human experts assign an integer score from 0 to 4 to each $\left<q, e, d\right>$ triplet.
The score represents whether the content of the document is off-topic(0), slightly relevant(1), relevant(2), useful(3), or vital(4) to the user search query and its potential intent, namely the event information.
Appendix ~\ref{appx:dataset_example} provides some examples from the testing data.

\vspace{-0.5em}
\subsubsection{Automatic Data Annotation}
\label{subsec:annotation}
The aforementioned samples collected from the production environment do not contain any relevance labels.
We apply an automatic process to annotate these unlabeled samples.
As illustrated in Figure~\ref{fig:data_arch},
our data annotation pipeline primarily consists of three steps:
1) A $\left<q, e, d\right>$ triplet collect from search logs is first split into a $\left<q, d\right>$ pair and a $\left<e, d\right>$ pair.
Meanwhile, the correlations between the query and its event are stored in two temporary dictionaries for subsequent data recovery. i.e. query-event dictionary and event-query dictionary.
2) Then, the two pairs are separately fed into our automatic annotation process for data annotation.
3) After obtaining the relevance label,
The labeled triplets are restored to quadruplet form by querying corresponding pre-cached dictionaries.
The first one is defined as the query-centric sample,  denote as $\left<q, e, d, r_{qd} \right>$, where the label $r_{qd}$ represents the relevance between query and document.
Similarly, The second one is expressed as $\left<q, e, d, r_{ed}\right>$ and called the event-centric sample, where the label $r_{ed}$ denotes the relevance between the event and document.

To minimize the data annotation costs, we designed a two-stage data annotation approach:

\textbf{Stage1: ModelZoo-based Coarse Annotation}.
In this step, large-scale unlabeled samples are input into a variety of existing matching models,
including BM25, Sentence-Bert~\citep{reimers2019sentence}, monoBERT~\citep{nogueira2019multi}, etc.
We refer to these multiple models as ModelZoo in this paper.
The majority voting algorithm~\citep{onan2016multiobjective} is applied to roughly classify the sample into either an easy or hard category:
1) When the majority of models vote consistently, the voting result exhibits a high degree of confidence, and the sample can be considered an easy sample, which is then directly added to the final dataset.
2) Otherwise, it is considered a hard sample and input to the LLMs for further discrimination.
Note that the prediction score of each model is a floating number, we use predefined human-experienced thresholds to map the model raw outputs into a binary category, i.e. positive class or negative class.

This approach allows for the swift annotation of large-scale unsupervised data.
However, it presents two critical issues:
1) The annotation granularity is overly broad, merely dividing samples into relevant and irrelevant categories. It fails to accommodate special scenarios such as \textit{weak relevance}, which are crucial in our industrial application contexts.
2) The accuracy of annotated data is generally low due to the limited generalization capabilities of existing models, thereby capping the potential performance of our retrieval model.
We will next adopt a more powerful model to carry out more accurate data labeling.

\begin{figure*}[ht]
    \centering
    \setlength{\abovecaptionskip}{0cm}  
    \includegraphics[scale=1]{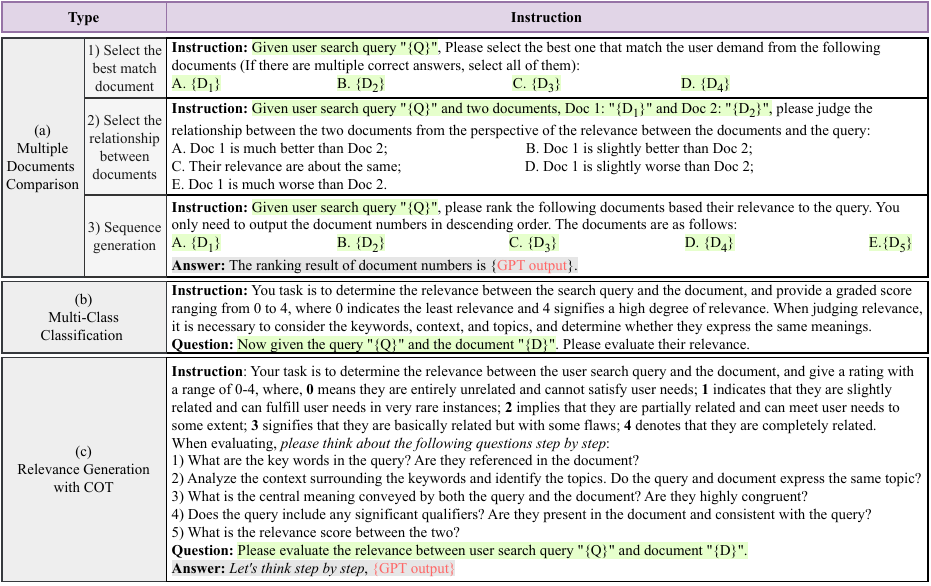}
    \caption{
        Different types of instructions for relevance annotation. The text highlighted in \colorbox{MyLightGreen}{light green} would change dynamically with different inputs, where $\lbrace{Q\rbrace}, \lbrace{D\rbrace}$
        are the placeholders of query and document, respectively.
    }
    \label{fig:llm_instruction}
\end{figure*}

\textbf{Stage 2: LLM-driven Fine Annotation}.
LLMs have demonstrated a remarkable ability to generalize zero-shot to various language-related tasks.
Therefore, we attempt to use LLM to annotate the difficult samples that are challenging for the aforementioned voting method.
In this section, we designed several different instructions for more precise data annotation.
The instructions are listed and depicted in figure ~\ref{fig:llm_instruction}.

\vspace{-1ex}
\begin{itemize}
\vspace{-1ex}
\item \textit{Multiple Documents Comparison}.
Since we adopt triplet loss to learn the partial ordering between two samples, obtaining an absolute label for each sample is not necessary.
Therefore, we designed instructions for comparing the relationships between documents.
As figure ~\ref{fig:llm_instruction}(a) shows, there are three instructions:
1) The first instructs the LLM to directly select the most relevant document corresponding to the query from various candidate documents;
2) The second instruction requires the LLM to compare the strength of the relevance relationship between two documents and a given query.
3) The third instruction ask the LLM to generate the permutation of documents in descending order based on their relevance to the query.
We believe that these designs can effectively and straightforwardly obtain pairwise training samples.

\item \textit{Multi-Class Classification}.
We divide the relevance between query and document into multiple levels, ranging from completely irrelevant to perfectly relevant.
Unlike the commonly-used multi-category classification, the class labels in our task incorporate information about relative ordering.
Furthermore, the number of classes is critical: having too few classes results in coarse targets that are not conducive to our application, while having too many classes leads to unclear distinctions between each class, particularly for adjacent classes.
Figure ~\ref{fig:llm_instruction}(b) is our instruction about multi-class classification.
Considering the practice of other works ~\citep{liu2021pre},
we set the number of classes to 5 to balance the difficulty of annotation and the effectiveness of the application.

\item \textit{Relevance Generation with CoT}.
Chain-of-Thought~(CoT) prompting enables LLMs to solve complex reasoning tasks by generating an explanation before the final prediction ~\citep{kim2023cotever}.
Based on the factors that human experts would consider during relevance evaluation and annotation,
the task is broken down into multiple steps, each of which considers the matching degree of different aspects,
such as whether the core words match, whether the topics match, whether the core semantics match, etc.
We prompt the LLM to \textit{think about specific questions step by step}, as figure ~\ref{fig:llm_instruction}(c) shows.
The generated results thereby would contain plausible explanations and the answers might be more precise.

\end{itemize}

The effects and experimental results of these instructions are compared in section ~\ref{sec:instructions}.
We choose the instruction that is most consistent with the labeling results of human experts for our fine-grained relevance annotation.


\subsection{Two-Stage Training Paradigm}
\label{sec:training_paradigm}
Due to constraints in search system performance, cost, and other factors, the majority of search engines can only recall a limited number of documents during the retrieval phase.
To enhance the retrieval performance of our model and achieve more effective recall of top relevant documents from billions of candidates,
inspired by previous work, such as \citet{liu2021pre}, Que2Search~\citep{liu2021que2search}, we designed a two-stage training paradigm for model training,
as shown in figure~\ref{fig:model_arch}(b).

\vspace{-0.5em}
\subsubsection{First-Stage Training}
In this stage, we use the large-scale business data annotated by ModelZoo to train a retrieval model that is suitable for real-time search scenarios.
Since the data annotated by ModelZoo is mostly of types that existing models can handle well and has similar data distribution,
to enhance the diversity of training data and improve training efficiency and effectiveness,
we adopt the following tricks to construct negative samples dynamically.

\textbf{Top-$k$ Hard Negative Sampling}.
Usually, negative data obtained through random sampling are easily distinguishable from positive data.
To solve this trouble,
for each query, we calculate its similarity score with each document and then sort them in descending order.
The document ranked $k$ is selected as the hard negative sample, where $k$ is a predefined hyper-parameter, usually greater than 1 to alleviate over-fitting.
The top-$k$ sampling method introduces more hard negatives and avoids overly easy negative samples,
thereby enhancing the robustness and diversity of the training data.
It is worth noting that due to the suboptimal retrieval performance of PTM, we initially apply random sampling.

\textbf{Cross Batch Negative Sampling}.
The effectiveness of in-batch negative sampling is inherently dependent on the size of the mini-batch.
Increasing the mini-batch size $N$ typically benefits negative sampling schemes and enhances performance, but it is often limited by GPU memory constraints.
In this paper, we employ a global memory bank to cache the document embeddings across the most recent $m$ mini-batches.
For each training batch, all positive documents in each pair are pushed into the buffer.
We then utilize the top-$k$ hard-negative sampling method mentioned previously to obtain hard-negative data and remove them from the buffer.
Note that the memory bank is updated with document embeddings, eliminating the need for any additional computation.

\subsubsection{Second-Stage Training}
After fine-tuning the large-scale data in the first stage,
the model has performed quite well on our business data.
However, the above model is trained based on binary classification data, and is difficult to distinguish subtle differences between different documents,
such as the critical \textit{weak relevance} case in industrial-level application scenarios.
Therefore, we further fine-tune our retrieval model produced by the first training stage on the LLM-annotated multi-class data,
which we consider to be more accurate and elaborate.


\section{Experiments}

\subsection{Evaluation Metrics}

\subsubsection{Metrics for Data Annotation}

\textbf{Cohen's Kappa}~\citep{cohen1960coefficient}
is a statistical coefficient that represents the degree of accuracy and reliability in statistical classification.
It measures the agreement between two raters who each classify $N$ items into $C$ mutually exclusive categories.
A higher kappa value indicates greater consistency in the annotation results of the two raters.

\subsubsection{Metrics for Offline Evaluation}
\label{metrics:offline}

We report various metrics on our human-labeled testing data for offline evaluation,
including recall@50, MAP@50, and MRR.
\textbf{Recall@$k$}~\citep{thakur2021beir} is a measure to evaluate how many correct documents are recalled at top-$k$ results.
\textbf{MAP@$k$}~\citep{enwiki:1146187267} is considered a reasonable evaluation measure for emphasizing returning more relevant documents earlier.
\textbf{MRR}~\citep{enwiki:1107032139} averages the reciprocal of the rank of the most relevant document over a set of queries. In this paper, we use the MRR metric to indicate the ranking of the first event-relevant document, with a higher MRR score signifying a higher position for the event-related document in the overall retrieval results.


\subsubsection{Metrics for Online Evaluation}\label{metrics:online}

\textbf{$\Delta$GSB}~\citep{zou2021pre} is a metric measured through side-by-side comparison.
For a user-issued query, the human experts are required to judge whether the new system or the base system gives better search results.
\textbf{CTR}~\citep{enwiki:1159144465} is the ratio of clicks on a search result page to the number of times a page is shown.
\textbf{DT}~\citep{enwiki:1158925169} stands for Dwelling Time, which measures the amount of time a user spends viewing a document after clicking a link from search results. An increase in this metric indicates that more search results are meeting the user's needs.
\textbf{QRR}, or Query Rewrite Rate, represents the percentage of users who modify their search queries while searching.
A high QRR indicates that users are unable to find satisfactory results and may need to refine their search terms several times.



\vspace{-1.5ex}
\subsection{Instructions Evaluation}
\label{sec:instructions}

\begin{figure}[t]
    \centering
    \includegraphics[scale=0.35]{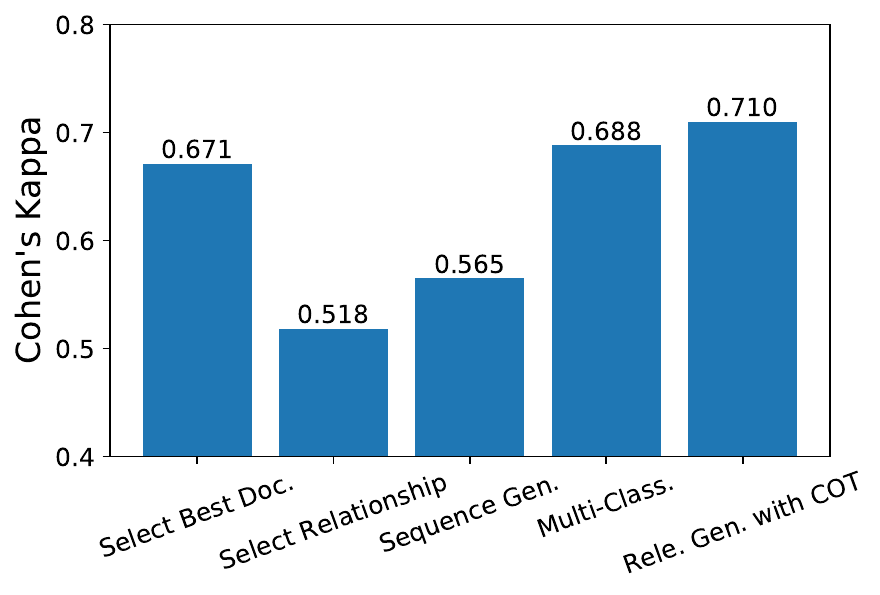}
    \vspace{-1em}
    \caption{The Consistency between Manual Data Annotation and LLM Data Annotation under Different Instructions.}
    \vspace{-3ex}
    \label{fig:instruction_kappa}
\end{figure}

To evaluate the effectiveness of various annotation tasks, we randomly sampled 1000 $\left<q, d\right>$ pairs and assigned them to experts on a crowdsourcing platform for manual annotation. Each pair was assigned a 0-4 grade based on relevance. These pairs will serve as a benchmark for different LLM labeling instructions.

Due to the diverse nature of annotation tasks, comparing the annotation results across different tasks poses a significant challenge in terms of achieving relative comparability.
Therefore, we standardized the results of different instructions into a document pair comparison format using the following methods:
1) For multi-class tasks, we converted the multi-class labeling results into a relative ranking format between two documents.
2) For document selection tasks, we considered the most relevant document identified by the LLM as the positive example, and the remaining candidates as negative samples. This was also transformed into a relative ranking format.
3) For sequence generation tasks, any two documents at different positions within the sequence were treated as positive and negative samples, forming document pairs.
By unifying human expert annotation results and LLM labeling results into a relative ranking format, we categorized the relationship between two documents as better(1), same(0), or worse(-1).

We used Cohen's Kappa metric to measure the consistency of annotation results. The experimental conclusions, presented in Figure~\ref{fig:instruction_kappa}, demonstrate that employing the \textit{Relevance Generation with COT} instruction yields highly consistent labels with human annotations. As a result, we adopt this instruction for our fine-grained automated data annotation.

\subsection{Baseline Comparison}
\begin{table}[t]
    \centering
    \small
    \setlength{\tabcolsep}{15pt} 
    \begin{tabular}{lccc}
        \toprule[1pt]
        { } & {Recall@50} & {MAP@50} & {MRR}  \\
        \hline
        ColBERTv2  & 0.8500 & 0.6217 & 0.8565 \\
        DPTDR & 0.8328 & 0.6087 & 0.8285 \\
        \textbf{ERR}   & \textbf{0.8552} & \textbf{0.6261} & \textbf{0.8956} \\
        \bottomrule[1pt]
    \end{tabular}
    \caption{The comparison between ERR and the baselines.}
    \vspace{-1.5em}
    \label{tab:baseline_comparsion}
\end{table}

In this section, to demonstrate the effectiveness of our proposed model,
we compared its performance with existing powerful retrieval models, such as  ColBERTv2~\citep{santhanam2021colbertv2} and DPTDR~\citep{tang2022dptdr}.
We fine-tuned these models on the same training data to eliminate data interference.
The offline evaluation metrics on our test dataset are shown in table~\ref{tab:baseline_comparsion},
and the result shows that ERR achieves the best performance on most of the metrics and surpasses the baseline models by a significant margin.
e.g. comparing with DPTDR model, ERR achieves nearly $2.4\%$, $1.7\%$ and $7.1\%$ improvements on the recall@50, MAP@50 and MRR metrics, respectively.
Compared to the baseline model, our model has demonstrated significant improvement in the MRR metric, which reflects the retrieval of event-related documents. This clearly highlights the effectiveness of our approach.

\subsection{Offline Ablation Study}
\begin{table}[t]
    \small
    \centering
    \setlength{\tabcolsep}{14pt} 
    \renewcommand{\arraystretch}{1} 
    \begin{tabular}{lccc}
        \toprule[1pt]
        ~ & Recall@50 & MAP@50 & MRR \\
        \hline
        \textbf{ERR} & \textbf{0.8552} & \textbf{0.6261} & \textbf{0.8956} \\
        \hline
        w/o-event & 0.8001 & 0.4711 & 0.8174 \\
        \hline
        w/o-CA & 0.8223 & 0.5345 & 0.79601 \\
        \hline
        w/o-CBS & 0.8253 & 0.4817 & 0.8500 \\
        w/o-THS & 0.8355 & 0.4911 & 0.8629 \\
        \hline
        w/o-UCL & 0.8346 & 0.4911 & 0.8663 \\
        w/o-ECT & 0.8191 & 0.4741 & 0.8422 \\
        \hline 
        w/o-TST & 0.8526 & 0.6190 & 0.8872 \\
        \bottomrule[1pt]
    \end{tabular}
    \caption{
    Ablation study on different components.
    }
    \vspace{-3em}
    \label{tab:ablation}
\end{table}

We study the effectiveness of each strategy by changing one strategy at a time.
As described in table~\ref{tab:ablation}, the validity of our model comes from the following components:

\vspace{-0.5ex}
\subsubsection{The effects of Event Info}
The event info is introduced to describe instant search intent and help recall the latest event-related documents.
To evaluate the influence of event information, we simultaneously removed the event input, event encoder, and cross-attention component.
Instead, we conducted the experiment solely utilizing the output of the query encoder as the query-side embedding.
The experimental results,
displayed in the third row of table ~\ref{tab:ablation}, clearly indicate that the model is generally less effective when only the query is used without events.

\vspace{-0.5ex}
\subsubsection{The effects of Cross-Attention}
ERR applies the cross-attention mechanism to fuse query and event fields so as to get a better trade-off.
w/o-$\operatorname{CA}$ implies the removal of cross-attention for the ERR model, concatenating the encoder outputs of the query and event directly.
The experimental results in table~\ref{tab:ablation} demonstrate that cross-attention plays an important role in data fusion — without which the model performance decrease on all of the metrics.

\vspace{-0.5ex}
\subsubsection{The effects of Negative Sampling}
We arrange two experiments in this part:
First of all, w/o-$\operatorname{CBS}$ indicates that we replace cross-batch sampling with in-batch sampling.
A significant decline in recall metrics can be observed from the experimental results.
This shows that our global ensemble sampling approach can increase the diversity of negative samples, which in turn improves the performance of the model.
Secondly, w/o-$\operatorname{THS}$ indicates removing the top-$k$ hard negative sampling strategy and employing random sampling instead.
We find that the model decreased dramatically in all recall metrics.
Top-$k$ hard sampling encourages the model actively learn more indistinguishable negative samples.

\vspace{-0.5ex}
\subsubsection{The effects of Multi-task Learning}
We bring unsupervised contrastive learning loss and two triplet losses with different objectives together for multi-task learning.
Firstly, w/o-$\operatorname{UCL}$ means removing the unsupervised contrastive learning loss and only using triplet losses, as compared to ERR.
Observing the training process, we find that unsupervised contrastive learning can speed up the convergence procedure in efficiency.
The ablation experiment $\operatorname{w/o-UCL}$ further proves that the unsupervised contrastive learning can improve the recall performance of the retrieval model to some extent compared with the direct usage of triplet losses.
Secondly, w/o-$\operatorname{ECT}$ means removing the event-centric task loss compared to ERR, through which we find that the recall metrics significantly decreased, which fully demonstrates the importance of the event-centric task for overall performance.

\vspace{-0.5ex}
\subsubsection{The effects of Two-stage Training}
To verify the effectiveness of the two-stage training, we mix and randomly shuffle the samples utilized in this training process,
and then train another model for the purpose of comparison.
It is evident that, when compared to ERR, the model trained solely in a single stage exhibits varying degrees of decline across different evaluation metrics.
This outcome serves as compelling evidence, underscoring the necessity of implementing the two-stage training approach.

\vspace{1ex}
The data above indicates that the experimental groups lacking event information and event-centric task loss exhibit the most significant decrease in evaluation metrics, indicating that the introduction of event information, and its enhanced utilization in the training process, have yielded significant retrieval performance gains. In addition, the implementation of other training strategies, such as the negative sampling strategy and unsupervised contrastive learning, has also positively impacted the results.

\vspace{-0.5ex}
\subsection{Online Evaluation}
\begin{table}[t]
    \vspace{-0.1em}
    \centering
    \small
    \setlength{\tabcolsep}{10pt} 
    \begin{tabular}{lcccc}
        \toprule[1pt]
        metric & $\Delta \rm{GSB}$ & CTR Gain & QRR Gain & DT Gain \\
        \hline
        ERR    & $+16.8\%$ & $+4.3\%$ & $-4.9\%$ & $+5.6\%$\\
        \bottomrule[1pt]
    \end{tabular}
    \caption{Online Experimental of ERR.}
    \vspace{-3em}
    \label{tab:online_experiments}
\end{table}

We have deployed ERR on our online search system and compared it with the existing base model.
By search expert annotation, ERR increases $\Delta \rm{GSB}$ metric by +$16.8\%$ on random queries.
After 6 consecutive days of online A/B testing, millions of user feedbacks indicate that ERR outperforms the baseline model in all metrics, and gains the average improvement of $+4.3\%$, $-4.9\%$ and $+5.6\%$ on CTR, QRR, and DT, respectively.
All of these experimental results prove that the proposed mechanisms bring substantial enhancements to the online search system.

\vspace{-0.5ex}
\section{Conclusion}
In this paper, we developed and deployed a real-time retrieval approach, namely ERR, for our news search business. ERR enhances retrieval performance by combining queries with breaking events related to the queries. Cross-attention and multi-task training was used to fuse events and queries. Additionally, we adopted a two-stage data annotation approach, consisting of a ModelZoo-based Coarse Annotation and an LLM-driven Fine Annotation, to obtain data for timely retrieval and reduce data annotation costs. Our proposed approach was extensively evaluated through offline experiments and online A/B tests, which demonstrated its effectiveness and usability.

\bibliographystyle{ACM-Reference-Format}
\bibliography{references}


\newpage

\appendix

\section{Dataset}
\subsection{Example of Queries in Real-time Search}
\begin{figure*}[h]
    \centering
    \setlength{\abovecaptionskip}{0cm}
    \includegraphics[scale=0.6]{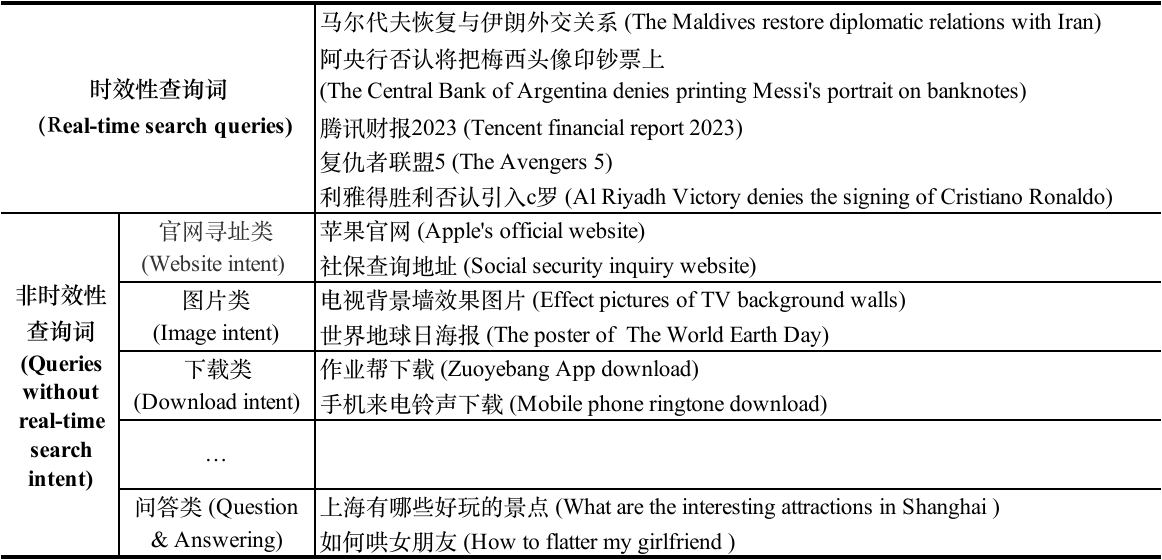}
    \vspace{0.5em}
    \caption{
        Queries from the production environment.
    }
    \label{fig:query_example}
    \vspace{2em}
\end{figure*}

Queries in universal search engines are diverse and hard to classify into fixed sets of categories.
We classify queries according to the user search intent, such as image intent, which means the user wants to find image-related resources; download intent represents the user's desire to find a download link for a movie, music, or App.
In a real-time search scenario, we simply divide the queries into two categories, namely real-time search queries and others.
Figure ~\ref{fig:query_example} shows the examples of queries and their types, all of which are derived from the search logs of our real-world production environment.

\subsection{Example of Testing Data}
\label{appx:dataset_example}
\begin{figure*}[ht]
    \centering
    \setlength{\abovecaptionskip}{0cm}
    \includegraphics[scale=0.8]{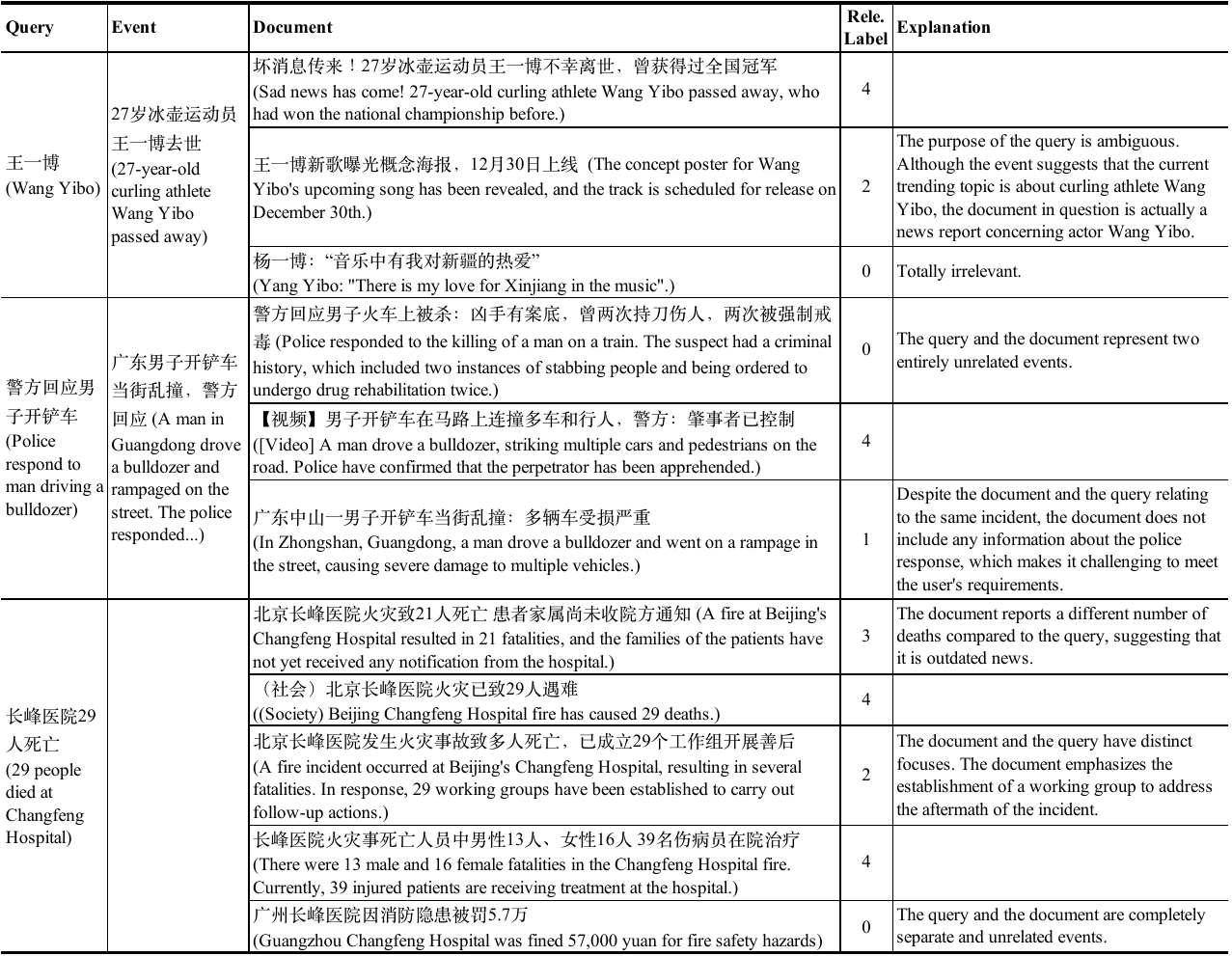}
    \vspace{0.5em}
    \caption{
        Some examples from our testing data.
    }
    \label{fig:dataset_example}
\end{figure*}

This section provides some examples of our testing data
(as shown in Figure ~\ref{fig:dataset_example}.
The table displays some classic scenarios in news search along with the standard relevance labels for them. The first two cases demonstrate that event information can effectively supplement unclear user needs. Each example is labeled with a score indicating the degree of relevance between the document and the query. The labels are explained as follows:
0: The document is completely unrelated to the query.
1: The document is related to the query but has a different focus, which does not meet the user's requirements.
2: The document is related to the query, but the query's purpose is ambiguous.
3: The document is related to the query, but the information in the document differs from that in the query.
4: The document is related to the query, and the information in the document matches the information in the query, meeting the user's requirements.

\subsection{Dataset Statistics}
\label{appx:dataset_stat}
The statistical analysis of both training data and testing data are shown in table \ref{tab:dataset_stat}.
The table provides an overview of the dataset size, including both training and testing data. The training data consists of tens of millions of query-document pairs, while the testing data contains 3,273 queries, 977 events, and 40,426 documents, resulting in a total of 128,281 query-document pairs.

\section{Implementation Details}
Here are the specific experimental details, including model implementation, training process, running platform, and data strategy.
1) The ModelZoo contains five models, which are BM25, Sentence-BERT, MonoBERT, ANCE and DPR, whose relevance thresholds are set to 4.3, 0.8, 0.75, 0.82, and 0.9, respectively.
For a pair of data, when the scores of at least 4 of these models reach their thresholds, we consider it as a high-confidence positive sample.
2) We utilize the Azure OpenAI Service\footnote{Azure OpenAI Service provides REST API access to OpenAI's powerful language models including the GPT-4 model. https://learn.microsoft.com/en-us/azure/cognitive-services/openai/overview} to employ the GPT-4 model for hard samples annotation.
3) Both query and document tower adopt RoBERTa-base as the encoder that contains 12 transformer layers with a dimension size of 768.
Documents, queries, and events are truncated to a maximum of 128 tokens, 24 tokens, and 36 tokens, respectively.
The output embedding of both query and document tower are compressed to 256 in dimension size.
Given the query-side and document-side embedding, we use cosine score as the similarity metric.
4) We train the model with Adam optimizer with 128 samples per batch.
The learning rate is set to $5e^{-5}$ with a linear warmup.
All hard negatives in each pair of samples are dynamically selected from a cross-batch global buffer with $8 \times batch$ in data size.
For multi-task training, The selection probability for the query-centric task is 0.7.
5) The model is implemented by the distributed PyTorch\citep{paszke2019pytorch} platform and trained on 8 NVIDIA Tesla A100 GPUs.
We further optimized ERR for accelerated inference using TensorRT library\citep{tensor_rt}.
The inference engine is deployed with FP16 computational kernels on a Tesla T4 GPU.

\begin{table}[t]
    \centering
    \setlength{\tabcolsep}{6pt}
    \begin{tabular}{lcccc}
        \toprule[1pt]
        {Dataset} & {\#Queries} & {\#Events} & {\#Docs} & {\#Q-D pairs}  \\
        \hline
        Training data & 4394798 & 1008619 & 3957072 & 64942852 \\
        Testing data  & 3273 & 977 & 40426 & 128281 \\
        \bottomrule[1pt]
    \end{tabular}
    \caption{The statistics of our dataset.}
    \label{tab:dataset_stat}
\end{table}



\end{document}